\documentclass[
aip,amsmath,amssymb, reprint, linenumbers
]{revtex4-1}

\usepackage{graphicx}
\usepackage{dcolumn}
\usepackage{bm}
\usepackage[ruled,vlined]{algorithm2e}
\usepackage[utf8]{inputenc}
\usepackage[T1]{fontenc}
\usepackage{mathptmx}
\usepackage{listings}

\usepackage{tikz}
\usepackage{soul}
\usetikzlibrary{arrows}
\usetikzlibrary{trees}
\usepackage{midfloat}
\usepackage{cancel}
\usetikzlibrary{decorations.pathmorphing}
\usetikzlibrary{decorations.markings}
\usetikzlibrary{automata,positioning}
\usepackage{float}
\usepackage{caption}
\usepackage{simplewick}
\usepackage{csquotes}
\usepackage{graphicx}
\usepackage{subcaption}
\usepackage{multirow}
\usepackage{amsmath}
\usepackage{cleveref}
\usepackage{tabularx}
\usepackage{natbib}
\usepackage{braket}
\usepackage{todonotes}
\usepackage{mathtools}
\presetkeys{todonotes}{inline,backgroundcolor=white}{}

\usepackage{xcolor}
\definecolor{codegreen}{rgb}{0,0.6,0}
\definecolor{codegray}{rgb}{0.5,0.5,0.5}
\definecolor{codepurple}{rgb}{0.58,0,0.82}
\definecolor{backcolour}{rgb}{0.95,0.95,0.92}
\lstdefinestyle{mystyle}{
    backgroundcolor=\color{backcolour},   
    commentstyle=\color{codegreen},
    keywordstyle=\color{magenta},
    numberstyle=\tiny\color{codegray},
    stringstyle=\color{codepurple},
    basicstyle=\ttfamily\footnotesize,
    breakatwhitespace=false,         
    breaklines=true,                 
    captionpos=b,                    
    keepspaces=true,                 
    numbers=left,                    
    numbersep=5pt,                  
    showspaces=false,                
    showstringspaces=false,
    showtabs=false,                  
    tabsize=2
}

\begin{document}

\nolinenumbers
\author{Dibyendu Mondal} \affiliation{Department of Chemistry, Indian Institute of Technology Bombay\\ Powai, Mumbai 400076, India} 
\author{Rahul Maitra}
\email{rmaitra@chem.iitb.ac.in} \affiliation{Department of Chemistry, Indian Institute of Technology Bombay\\ Powai, Mumbai 400076, India} 
\affiliation{Centre of Excellence in Quantum Information, Computing, Science \& Technology, \\ Indian Institute of Technology Bombay, \\ Powai, Mumbai 400076, India}

\title{Ground or Excited State: a State-Specific Variational Quantum Eigensolver for Them All}

\begin{abstract}
    Variational Quantum Eigensolver (VQE) provides a lucrative platform to determine molecular
    energetics in near-term quantum devices. While the VQE is traditionally tailored to 
    determine the ground state wavefunction with the underlying Rayleigh-Ritz principle, 
    the access to specific symmetry-adapted excited states 
    remains elusive. This often requires high depth circuit
    or additional ancilla qubits along with
    prior knowledge of the ground state wavefunction. We propose a unified VQE framework that treats the ground
    and excited states in the same footings. 
    With the knowledge of the irreducible representations of the spinorbitals, we construct a 
    multi-determinantal reference that is adapted to a given spatial symmetry where additionally, the determinants are entangled through appropriate 
    Clebsch-Gordan coefficients to ensure the desired spin-multiplicity. We introduce the notion of 
    totally symmetric, spin-scalar unitary which maintains 
    the purity of the reference at each step of the
    optimization. The state-selectivity safeguards the 
    method against any variational collapse while leading 
    to any targeted low-lying eigenroot of arbitrary symmetry. 
    The direct access to the excited states shields our 
    approach from the cumulative error that plagues excited 
    state calculations in a quantum computer and with 
    few parameter count, it is expected to be realized in near-term quantum devices. 
\end{abstract}

\maketitle

\date{\today}

\section{Introduction}

The determination of molecular energy and properties is one of the most 
promising applications of quantum computing in the near-term quantum computers.
The exploration of atomic and molecular many-body physics and chemistry
is often restricted by the exponential growth of the Hilbert space\cite{expo1,expo2}
which makes it intractable in a classical computer. In spite of the 
limitations in the existing quantum architecture that often suffers from 
their limited coherence time and poor gate fidelity, efforts have 
been made to design various algorithms to determine the ground state 
energetics that can be realized in quantum hardware architecture. 
As such, the variational Quantum Eigensolver (VQE)\cite{Peruzzo_2014,Tilly_2022,ucc_review} is widely considered to 
be realizable in the noisy intermediate scale quantum (NISQ) devices and 
has actually been implemented in various hardware platforms\cite{Peruzzo_2014,Trapped_ion,quantum_pro} to find the 
"best" variationally optimized wavefunction for the ground state in 
terms of a number of parameters.

Along with the ground state, the access to the molecular excited states is crucial to 
simulate chemistry. For a plethora of processes in nature, 
like photosynthesis, photochemical reactions as well as synthetic chemistry like the 
functioning of solar cells require an accurate description of excited states. Thus, 
quantum algorithms should be tailored accordingly such that the calculations involving
excited states can exploit the quantum supremacy. 
Historically, quantum phase estimation (QPE)\cite{QPE} was developed as the most versatile
method that treats ground and excited states in the same footing\cite{esqpe1,esqpe2,esqpe3,esqpe4}. However, QPE 
necessitates a sufficiently long coherence time over extremely deep and complex 
quantum circuit for which a large number of fault-tolerant 
qubits are required. It is thus supremely important to develop quantum algorithms 
within the near-term VQE framework\cite{Peruzzo_2014} that may treat excited states with a reasonable
accuracy, overcoming the detrimental effects of quantum noise.

There exists a number of methods that access to the excited states in terms of 
excitation energy with respect to the ground state. Within the unitary coupled cluster (UCC) parametrization of the 
unitary, these conceptually carry the 
signature of classical many-body theories to treat quantities of spectroscopic 
energy difference via the diagonalization of an effective Hamiltonian in a truncated fermionic 
subspace. For example, the Quantum Subspace Expansion (QSE)\cite{QSE1,QSE2,QSE3,QSE4} is akin to the configuration
interaction (CI) in classical many-body theory and is built upon a prior knowledge of 
the correlated ground state wavefunction. A vexing issue that often plagues its applicability is
its lack of size extensivity and thus size-consistency. The quantum equation of motion (qEOM)\cite{qEOM}
is often considered a viable alternative, although their choice of the operator manifold 
is somewhat physically ambiguous in the absence of a 'killer condition'. This problem was
alleviated by introducing a self-consistent operator manifold in the EOM framework\cite{eom_vqe} 
that was shown to recover the killer condition and size-
extensivity in its description of the spectroscopic
quantities.

There exist a suite of complementary approaches where 
one directly determines the excited states by optimizing 
a cost function in a step-wise manner. For example, in 
the orthogonal state reduction variational eigensolver
(OSRVE) algorithm\cite{OSRVE}, the determination 
of the $k-$th excited root necessitates one to additionally 
have $k$ sets of optimally parametrized circuits, as well as 
$k$ auxiliary qubits to eliminate the contribution from 
lower $(k-1)$ states, including the ground state. One may 
of course, design an effective
hamiltonian, parametrized with an optimally chosen constant, 
in such a way that the excited states are orthogonal 
to all other lower energy states (which include the 
ground state as well). The variational quantum deflation
(VQD) algorithm\cite{VQD} is based upon this idea. 
One may impose additional parametrization to automatically
adjust the constant to define the effective Hamiltonian. 
The spin-symmetry of the state may get somewhat restored via
an additional constraint of spin projection in the effective
hamiltonian\cite{VQD_spin}. This, of course, hinges upon 
the availability of additional quantum resources including 
a high gate depth implementation, high number of quantum 
measurements and auxiliary qubits. 

In addition to the aforementioned approaches, there exist 
alternative techniques that are employed to minimize the 
energy variance or von Neumann entropy or to find the 
expectation values of the squared Hamiltonian for the 
purpose of characterizing the higher-energy states. 
The principle of the variance-VQE\cite{variancevqe} 
method is to minimize the energy variance, which will 
be minimum only for eigenstates of the Hamiltonian.
Conversely, the Folded Spectrum Method\cite{foldedspectrum}
necessitates the determination of the expectation
value of the squared Hamiltonian operator, entailing 
an expansion with a quadratic increase in terms. 
However, the methodology known as the Witness Assisted
Variational Eigensolver Spectra (WAVES)\cite{WAVES}
embarks on a dual optimization trajectory: minimization 
of the ansatz energy expectation as well as the von 
Neumann entropy of a control ancilla qubits, which however, 
requires a large number of high-depth controlled unitaries.     

Apart from the high implementation cost of these methods,
one common drawback of them all is that the the notion of
the molecular symmetry (spatial and spin) 
elements remain extremely fuzzy. One may note that any
information related to the excited state or the
spectroscopic energy difference hinges upon the specific
knowledge of the symmetry elements of the excitations/excited
states. It is thus imperative that the information of the
state symmetry is restored. We try to address all these
issues in one go with a simple concept that seems to have
profound implications: \textit{in the conventional UCC
framework, if the evolution operator is 
designed in a way that is both spatial and spin 
symmetry-scalar, the variational optimization of an 
arbitrary energy functional will generate the correlated
state which has the exactly same symmetry elements as 
the reference state.} Towards this, we develop
state-specific VQE (ss-VQE, to distinguish from 
Subspace-Search VQE (SSVQE)\cite{SSVQE}) that addresses 
one root at a time, characterized by specific symmetry
elements. This 
is achieved by the construction of a desired symmetry 
adapted reference, followed by the evolution with 
a symmetry scalar unitary in an unconstrained manner 
such that the purity of the symmetry elements in the
reference is maintained. Our ss-VQE has clear multifold
advantage over all the existing methods:
\begin{itemize}
    \item For the first time, we introduce the notion of molecular 
    symmetry in an excited state calculation in quantum architecture which remained fuzzy so far. Thus this approach is expected to be supremely significant to spectroscopists. 
    \item Unlike \textit{all} other methods, ss-VQE does not 
    require any prior knowledge of the ground state and it can selectively target any low-lying excited state (with arbitrary spin-multiplicity and irreducible representation, as dictated by the symmetry of the molecule) to find the solution using regular VQE algorithm. 
    \item The optimization requires \textit{no additional
    quantum resources} than the ground state calculation. Furthermore, due to the direct determination of the excited root, it is safeguarded from the detrimental cumulative effects of noise that otherwise make excited
    state calculations highly inaccurate.
    \item For any excited state calculation based on ground
    state wavefunction, it is imperative that both the 
    states \textit{must} be parametrized in an equal 
    footing. The existing methods have so far used 
    independent parametrization for the ground and the
    excited states. While this may have its pros, one 
    however, can parametrize them in a way that artificially
    makes the excitation energy/excited state more accurate. 
    Our approach is unified for ground and excited states 
    and hence the excited state is as good as the ground 
    state, as it should be. 
\end{itemize}    

The rest of the manuscript is organized as follows: we start
with the introduction of the traditional VQE algorithm and
demonstrate how the conventional UCC approach
generates symmetry-contaminated solution while navigating 
to the ground state. In the subsection \ref{ref_pre.}, we demonstrate the
construction of symmetry-adapted multi-determinantal reference and
how various 
determinants are entangled via Clebsch-Gordan coefficients to 
construct the spin-adapted reference. In subsection \ref{unitary_evo.}, we define
the unitary that is spin-scalar and show certain cases how it is achieved by restricting the number of free parameters. Section \ref{results} will deal with the potential energy profile and 
dipole moment surface for a number of excited 
roots for several challenging systems that justify the superiority of the approach. Finally, we conclude in 
section \ref{conclusion}.

\section{Theory}
\subsection{VQE Algorithm for the Ground State and a Road-map to its Extension for the Excited States:}
One of the central problems of physics and chemistry is to find the exact ground and 
excited states of many-body systems.  
The variational quantum eigensolver is a hybrid near-term quantum-classical algorithm
that is tailored to find the upper bound of ground-state energy associated with a 
given Hamiltonian. The VQE algorithm may be considered as an application of time-independent 
Rayleigh-Ritz variational principle where a parametrized energy functional is expressed as:

\begin{eqnarray}
E(\theta) &=& \frac{\langle \Psi(\theta)|H|\Psi(\theta) \rangle}{\langle \Psi(\theta)|\Psi(\theta) \rangle} \nonumber \\
&=& \langle \Phi_{0}|U(\theta)^\dag HU(\theta)|\Phi_{0} \rangle, \label{vqe_energy} 
\end{eqnarray}

Here $|\Psi(\theta)\rangle$ is the parametrized trial wavefunction, and $H$ is the 
molecular Hamiltonian that can be represented in the second quantization as:
\begin{equation} \label{Hamiltonian}
\hat{H}=\sum_{p,q} h_{pq} a^{\dagger}_{p} a_{q}+\frac{1}{2}\sum_{p,q,r,s} h_{pqrs} a^{\dagger}_{p} a^{\dagger}_{q} a_{r} a_{s}
\end{equation}
The indices $p, q, r, s, ..$ are generic orbitals indices that are agnostic to the occupancy of the
given orbital. $h_{pq}$ and $h_{pqrs}$ are the one and two-electron integrals. 
$U(\theta)$ may be considered as a unitary evolution operator, parametrized by a set of
parameters $\theta$, that acts upon a suitably chosen reference state: $|\Psi(\theta)\rangle = U(\theta)|\Phi_0\rangle$. For the closed-shell 
systems, the reference state is often taken to be the Hartree-Fock determinant; however, one 
may choose to start with a different set of reference functions.   
The VQE algorithm may be divided into two subparts: firstly, it uses quantum architecture
to prepare an optimally parametrized state and the measurement of the associated 
energy expectation value. The information in the quantum subpart is then converted into 
classical data and fed into the classical routine. In the classical routine, the energy 
is minimized with respect to the parameters $\theta$ and the parameters are updated. 
Being a direct descendant of the Rayleigh-Ritz variational principle, the predicted 
energy provides the upper bound to the exact ground state energy.

\begin{eqnarray}
E_{0} = \mathrm{min}_{\theta} \langle \Phi_{0} | U(\theta)^\dag H U(\theta) | \Phi_{0} \rangle, \label{eqE1}
\end{eqnarray}

In often cases, the quality of the final state is often dictated by the 
choice of the parametrized ansatz which should be expressive to span a given
$N-$electron Hilbert space, yet at the same time the resulting circuit depth 
should not be beyond the limit of near-term NISQ realization. In this regard,
the unitary coupled cluster (UCC) provides a good starting point for a
parametrized ansatz in terms of various (anti-hermitian) cluster operators.
Thus with a choice of excitation operators $T$, the unitary $U(\theta)$ may 
be written as:
\begin{eqnarray}
  U(\theta)=e^{\tau(\theta)} ; \tau = T-T^{\dagger} 
  \label{Psitheta}
\end{eqnarray}
Here the cluster operator $T$ may be expressed as a sum of various many-body terms:
$T=T_{1}+T_{2}+T_{3}+...$ that can further be written in second quantized form as, 
\begin{eqnarray}\label{SD_cluster_operator}
T_{1}=\sum_{i,a} \theta^{a}_{i} a^{\dagger}_{a} a_{i}   \hspace{0.3cm} 
T_{2}=\sum_{i,j,a,b} \theta^{ab}_{ij} a^{\dagger}_{a} a^{\dagger}_{b} a_{j} a_{i} \hspace{0.2cm} \dots
\end{eqnarray}
In often cases and in all the subsequent developments in this manuscript, we truncate 
the excitation rank to two.

While great progresses have already been made in the VQE algorithm with UCC ansatz, 
there is little effort to date to encode the spatial and spin symmetry elements 
in the unitary operator or in the chosen reference function. Orbital symmetry 
requires a subtle understanding of 
group theory. Various symmetric molecules may be classified into different point 
groups and each such point group has various irreducible representations. For 
single excitations $T_i^a$, the orbital symmetry of $\phi_i$ must be same as that of 
$\phi_a$. Thus, if the symmetry operator $\hat{s_e}$ maps a given 
orbital to its irreducible representation $\hat{s_e}(\phi_i)=\sigma_i$,
then the symmetry requirement demands $\sigma_i = \sigma_a$ such that the 
direct product of the symmetry elements belongs to the totally symmetric representation\cite{symvqe}. 
Note that for the spinorbitals that share the same spatial function have
the same irreducible representations: $\hat{s_e}(\phi_p)=\hat{s_e}(\bar{\phi}_p)=\sigma_p$.
In addition, 
for the double (like $T_{ij}^{ab}$) and higher order excitations, the product 
of the irreducible representations of the occupied orbitals must be same as that of
the virtual orbitals: $\sigma_i\otimes \sigma_j = \sigma_a\otimes \sigma_b$. Considering only the Abelian point groups, the direct product of 
the product of the irreducible representations of all the occupied orbitals characterizing 
a given cluster operator and the product of all the virtual orbitals must belong to 
the totally symmetric representation. This
further implies that the unitary operator, given by the exponential of the totally symmetric
(anti-hermitian) cluster operator is also symmetry scalar. This translates to the fact
that if the reference determinant $|\Phi_0\rangle$ is taken to be adapted to a 
particular symmetry, the evolved state also should maintain the same symmetry:
\begin{equation}\label{symmetry_combination}
|\Psi^\Sigma\rangle = e^{\tau}\ket{\Phi_{0}^\Sigma}\rightarrow c_0|\Phi_{0}^\Sigma\rangle + \sum c_I|\Phi_I^\Sigma\rangle
\end{equation}
Here, superscript $\Sigma$ denotes the symmetry of the reference (and hence the correlated) 
state. The equation \ref{symmetry_combination} constitutes a comprehensive representation of the ground-state wave function for any inherent symmetry within a molecular system.

While the spatial symmetry of the target wavefunction is ensured by default when the 
cluster operators
are chosen to be totally symmetric, the spin symmetry is somewhat difficult to maintain. The 
cluster operators with an unrestricted inclusion of all the excitation terms with complementary spin 
mixes up various spin channels. For example, a two-body cluster operator may be written
as a sum of a singlet paired channel having a symmetric combination of opposite spin 
excitations and a triplet paired channel with an antisymmetric combination of 
opposite spin excitations plus all the same spin excitations\cite{oovqe,sing-trip}. Due to the nonlinearity
in the unitary operator, the various spin channels get mixed up during the unitary evolution.
This means if one prepares a reference state which is triplet and adapted to a given 
spatial symmetry, an unrestricted evolution through the exponential of the symmetry 
adapted (anti-hermitian) cluster operators
will generate a state of the same spatial symmetry, but a broken spin symmetry. This 
means if there is a low-lying singlet state (of that given spatial symmetry) available, 
this evolution, although started with a triplet reference, will land up in the 
lower energy singlet. One, however, may demand to eliminate the spin-dependence of 
the excitations by restricting their independent variation and appropriate subgrouping. 
The resulting cluster operators may be called as spin-scalar and thus any evolution with 
such spin-scalar operators would not lead to any undesired spin-contaminated solution. Below, we discuss how to construct a 
(spatial and spin-) symmetry-adapted reference, followed by the 
spin-scalar unitary towards the formulation of the unified
state-specific VQE framework.

\subsection{Development of the State-Specific Variational Quantum Eigensolver:}
\subsubsection{Symmetry Adapted Reference State Preparation: Theoretical Aspects and 
Quantum Circuit Design}\label{ref_pre.}
As discussed previously, we emphasize on preparing a reference state that is adapted to 
a given spatial and spin symmetry. We will first discuss the general working principle 
towards the construction of the symmetry-adapted determinant, followed by a particular 
example. One may note that the HF determinant where all the
orbitals are doubly occupied necessarily belongs to the totally symmetric representation
($\underline{A} = A$ or $A_1$ or $A_{1g}$). Thus if the spatial occupied orbitals (in HF determinant) 
are denoted by $i, j, k,...,m$, one may write: 
\begin{eqnarray}
\hat{s_e}(\phi_i)\otimes \hat{s_e}(\bar{\phi}_i)\otimes \hat{s_e}(\phi_j)\otimes \hat{s_e}(\bar{\phi}_j)\otimes \cdot \cdot \cdot \hat{s_e}(\phi_m)\otimes \hat{s_e}(\bar{\phi}_m)= \nonumber \\
\Big(\sigma_i \otimes \sigma_i\Big) \otimes \Big(\sigma_j \otimes \sigma_j\Big) \cdot \cdot \cdot \otimes \Big(\sigma_m \otimes \sigma_m\Big) = \underline{A}
\label{HF_symm}
\end{eqnarray}
Here the overbar denotes the down spinorbitals. Furthermore, 
note that the HF reference is necessarily coupled to a spin singlet. Starting from 
the HF reference, we now proceed to generate a reference state which is characterized
by the spatial symmetry $\Sigma$. Towards this, one may choose an unoccupied spatial
orbital, $\phi_e$ (with respect to the HF determinant) such that the direct product of the 
irreducible representations of $\phi_e$ and that of an high lying occupied orbital $\phi_m$ is
$\Sigma$:
\begin{eqnarray}
    \hat{s_e}(\phi_m)\otimes \hat{s_e}(\phi_e) = 
    \sigma_m \otimes \sigma_e = \Sigma
    \label{symadapt}
\end{eqnarray}
This translates to the fact that one may generate two complementary determinants
$\phi_A$ and $\phi_B$ which belong to the irreducible representation $\Sigma$.

{\small{
\begin{eqnarray}
    \hat{s_e}(\phi_A)=\hat{s_e}(\phi_i)\otimes \hat{s_e}(\bar{\phi}_i)\otimes \hat{s_e}(\phi_j)\otimes \hat{s_e}(\bar{\phi}_j)\otimes \cdot \cdot \cdot\hat{s_e}(\bar{\phi}_m)\otimes \hat{s_e}({\phi}_e) \nonumber \\
    =\Big(\sigma_i \otimes \sigma_i\Big) \otimes \Big(\sigma_j \otimes \sigma_j\Big) \otimes \cdot \cdot \cdot \Big(\sigma_m \otimes \sigma_e\Big) = \Sigma \nonumber \\
    \hat{s_e}(\phi_B)=\hat{s_e}(\phi_i)\otimes \hat{s_e}(\bar{\phi}_i)\otimes \hat{s_e}(\phi_j)\otimes \hat{s_e}(\bar{\phi}_j)\otimes \cdot \cdot \cdot\hat{s_e}({\phi}_m)\otimes \hat{s_e}(\bar{\phi}_e)  \\
    =\Big(\sigma_i \otimes \sigma_i\Big) \otimes \Big(\sigma_j \otimes \sigma_j\Big) \otimes \cdot \cdot \cdot \Big(\sigma_m \otimes \sigma_e\Big) = \Sigma \nonumber    
    \label{symadapt_md}
\end{eqnarray}}}

Since $\phi_A$ and $\phi_B$ are both adapted to a particular spatial symmetry $\Sigma$, 
their linear combination would also have the same irreducible representation. Furthermore, 
$\phi_A$ and $\phi_B$ are only different by a spin-complimentary excitation and hence
we will take their equally weighted linear combination 
$\phi_0^\Sigma = \frac{1}{\sqrt{2}}(\phi_A^\Sigma \pm \phi_B^\Sigma)$ as the reference determinant.
One may note that the state $\phi_0^{\Sigma}$ renders to 
be a multi-determinantal reference function rather than 
a single determinant and hence the spin-scalar parametrization
of the unitary would be rather vexed. Even more 
interestingly, there are further subtler aspects to the
choice of the multi-determinantal function: the choice of 
the relative phase would play a crucial role in deciphering
the singlet or the triplet spin-symmetry of the reference
determinant and we will discuss it in the 
subsequent paragraphs.

One may note, the spatial symmetry adaptation relies on 
finding a set of orbitals $\phi_m$ and $\phi_e$ for which Eq. 
\ref{symadapt} holds. The resulting Eq. \ref{symadapt_md}
then implies that the multi-determinantal state 
$\phi_0^{\Sigma}$ consists of two open shell 
configurations $\phi_A^\Sigma$ and $\phi_B^\Sigma$, each of 
which has two unpaired electrons. With all other orbitals 
(except $\phi_m$ and $\phi_e$) are doubly occupied, the overall 
spin multiplicity of the reference will be determined by 
the relative sign through which the singly occupied 
orbitals $\phi_m$ and $\phi_e$ are coupled to generate an overall
fermionic antisymmetric wavefunction. 

It is somewhat trivial to note that the multi-determinantal singlet and 
triplet state $\phi_0^{\Sigma}$ may trivially be written in terms of various
Slater determinants through the Clebsch-Gordan coefficients. 
While the singlet $\phi_0^{\Sigma}$ requires a symmetric combination 
of the spatial wavefunction, the $M_s=0$ triplet state takes the antisymmetric 
combination:
\begin{eqnarray}
    \phi_0^{\Sigma, S=0}=\frac{1}{\sqrt{2}}(\phi_A^{\Sigma}+\phi_B^{\Sigma}) \nonumber\\
    \phi_0^{\Sigma, S=1}=\frac{1}{\sqrt{2}}(\phi_A^{\Sigma}-\phi_B^{\Sigma})
\end{eqnarray}
where $\phi_A^{\Sigma}$ and $\phi_B^{\Sigma}$ can be represented as
{\small{\[
\phi_A^{\Sigma} = 
\begin{vmatrix}
    \phi_i^{\sigma_i}(1) & \bar{\phi}_i^{\sigma_i}(1) &  \phi_j^{\sigma_j}(1) & \bar{\phi}_j^{\sigma_j}(1) & ... & \phi_e^{\sigma_e}(1) & \bar{\phi}_m^{\sigma_m}(1)\\
    \phi_i^{\sigma_i}(2) & \bar{\phi}_i^{\sigma_i}(2) &  \phi_j^{\sigma_j}(2) & \bar{\phi}_j^{\sigma_j}(2) & ... & \phi_e^{\sigma_e}(2) & \bar{\phi}_m^{\sigma_m}(2)\\
    . & . & . & . & ... & . & .\\
    . & . & .& . & ... & . & .\\
    . & . & .& . & ... & . & .\\
    \phi_i^{\sigma_i}(N) & \bar{\phi}_i^{\sigma_i}(N) &  \phi_j^{\sigma_j}(N) & \bar{\phi}_j^{\sigma_j}(N) & ... & \phi_e^{\sigma_e}(N) & \bar{\phi}_m^{\sigma_m}(N)
    
\end{vmatrix}
\]}}
and
{\small{\[ \phi_B^{\Sigma} = 
\begin{vmatrix}
    \phi_i^{\sigma_i}(1) & \bar{\phi}_i^{\sigma_i}(1) &  \phi_j^{\sigma_j}(1) & \bar{\phi}_j^{\sigma_j}(1) & ... & \phi_m^{\sigma_m}(1) & \bar{\phi}_e^{\sigma_e}(1)\\
    \phi_i^{\sigma_i}(2) & \bar{\phi}_i^{\sigma_i}(2) &  \phi_j^{\sigma_j}(2) & \bar{\phi}_j^{\sigma_j}(2) & ... & \phi_m^{\sigma_m}(2) & \bar{\phi}_e^{\sigma_e}(2)\\
    . & . & . & . & ... & . & .\\
    . & . & .& . & ... & . & .\\
    . & . & .& . & ... & . & .\\
    \phi_i^{\sigma_i}(N) & \bar{\phi}_i^{\sigma_i}(N) &  \phi_j^{\sigma_j}(N) & \bar{\phi}_j^{\sigma_j}(N) & ... & \phi_m^{\sigma_m}(N) & \bar{\phi}_e^{\sigma_e}(N)
    
\end{vmatrix}
\]}}
Note that both $\phi_0^{\Sigma, S=0}$ and $\phi_0^{\Sigma, S=1}$ are adapted to 
the desired spatial symmetry $\Sigma$.

While the construction of the singlet or the triplet states is straightforward,
the adaptation of the reference state to other multiplets (even for the systems
with an even number of electrons) is somewhat convoluted and as such it is 
not explored in the present manuscript. One may build-up a Serber-like spin
coupling scheme via Clebsch-Gordan coefficients to adapt the reference 
functions to higher spin-multiplets and this will be the subject to a 
separate forthcoming publication.

\begin{figure*}[t]
    \centering
    \includegraphics[width= 17.25cm, height=7cm]{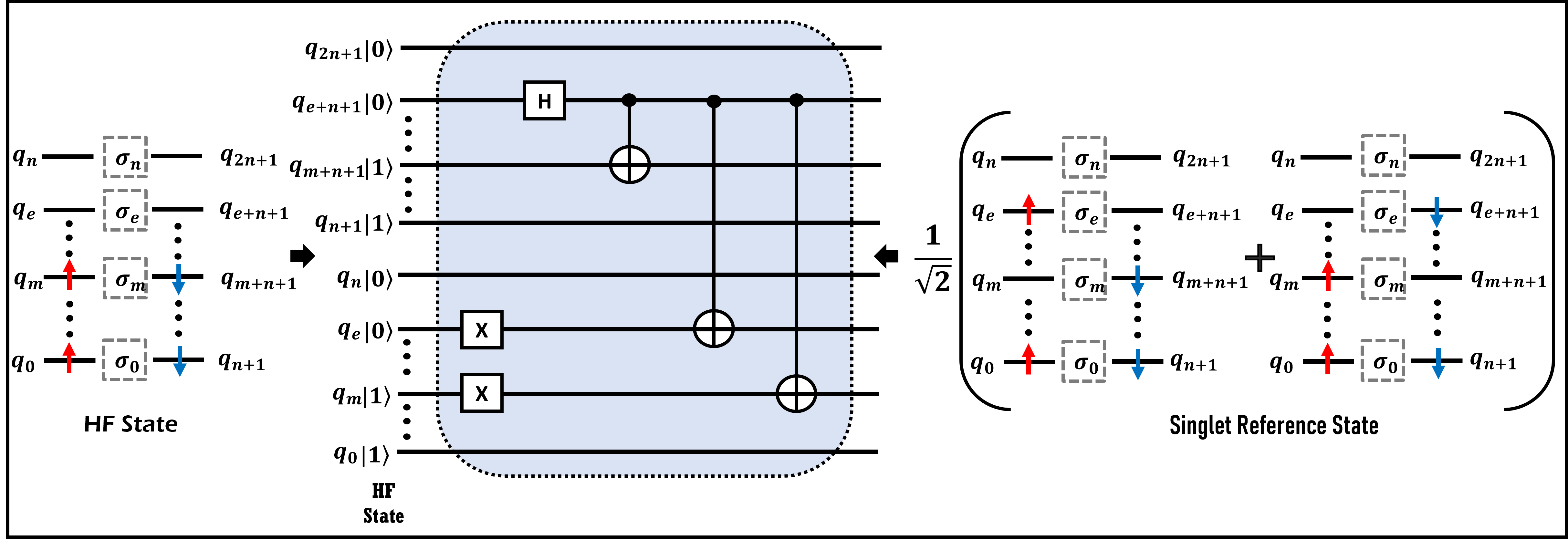}
    \caption{\textbf{Schematic quantum circuit to generate the singlet coupled reference with 
    spatial symmetry adapted to $\Sigma=\sigma_m\otimes\sigma_e$.}}
    \label{fig:circuit1}
\end{figure*}
Operationally, one may design a unitary transformation in terms of (non-parametrized) quantum circuit 
that generates $\phi_0^\Sigma$ starting from the HF determinant $\Phi_{HF}$: 
\begin{equation}
    |\phi_0^{\Sigma,S}\rangle = U^{\Sigma, S}_{A,S=0}|\phi_{HF}^{A,S=0}\rangle
\end{equation}
The superscript $S$ stands for the spin-multiplicity of the reference state $\phi_0$ with 
spatial symmetry $\Sigma$. Below we discuss the details of the circuit implementation
that performs the transformation.

In order to explain the intricacies of the circuit that 
constructs the multi-determinantal reference state, we take
a model system having $(n+1)$ spatial orbitals ($(2n+2)$ 
spin orbitals). Let the orbitals have the irreducible 
representation $\sigma_0, ...,\sigma_m, ...,\sigma_e,\sigma_n$. We have employed a direct mapping
between the spinorbitals and qubits, which means that 
$(2n+2)$ spin orbitals are mapped onto $(2n+2)$ qubits:
$q_0,...,q_m,...,q_e,q_n,q_{n+1},...,q_{m+n+1},...,q_{e+n+1},q_{2n+1}$. In our example, we assume that the first $(n+1)$
qubits represent up-spinorbitals, while the next $(n+1)$
qubits represent the down-spinorbitals. Thus the qubit 
representation for the HF state would be 
$\ket{0_{2n+1}0_{e+n+1}...1_{m+n+1}...1_{n+1}0_n0_e...1_m...1_0}$, where $0$ and $1$ stand for the occupation numbers 
for the corresponding spinorbitals and the subscripts refer to their position where the 
spinorbitals are arranged from right to left inside the ket. 
One may note that $\sigma_p=\sigma_{p+n+1}$ as both $\phi_p$ and $\phi_{p+n+1}$
share the same spatial orbital. With this, it is trivial to
show that the closed shell HF determinant belongs to the 
totally symmetric representation (Eq. \ref{HF_symm}).

Assuming the target symmetry of the reference function $\Sigma$ is a direct
product of $\sigma_m$ (where $\phi_m$ is a 
\textit{high-lying} occupied orbital in the HF reference) 
and $\sigma_e$ (where $\phi_e$ is a low-lying unoccupied orbital in the 
HF reference): $\Sigma=\sigma_m\otimes \sigma_e$. So as per Eq. \ref{symadapt_md}, 
the determinants that contribute to the reference function of symmetry $\Sigma$ are $\phi_A^\Sigma = \ket{0_{2n+1}0_{e+n+1}...1_{m+n+1}...1_{n+1}0_n1_e...0_m...1_0}$ and
$\phi_B^\Sigma = \ket{0_{2n+1}1_{e+n+1}...0_{m+n+1}...1_{n+1}0_n0_e...1_m...1_0}$. 
We will take equally weighted linear combinations of $\phi_A^\Sigma$ and 
$\phi_B^\Sigma$ to generate the singlet (symmetric combination) and 
the triplet (anti-symmetric combination) functions. Starting from the 
HF determinant, the circuit that generates the singlet function is shown 
in Fig:\ref{fig:circuit1} and the sequence of the gates that takes us
from the HF determinant to the singlet coupled entangled multi-determinantal
reference function is described below:
\begin{multline}
\ket{0_{2n+1}0_{e+n+1}...1_{m+n+1}...1_{n+1}0_n0_e...1_m...1_0}\xrightarrow[X(q_{m})]{X(q_{e})} \\
\ket{0_{2n+1}0_{e+n+1}...1_{m+n+1}...1_{n+1}0_n1_e...0_m...1_0} \xrightarrow[]{H(q_{e+n+1})} \\
\frac{1}{\sqrt{2}}\Big(\ket{0_{2n+1}0_{e+n+1}...1_{m+n+1}...1_{n+1}0_n1_e...0_m...1_0}+\\
\ket{0_{2n+1}1_{e+n+1}...1_{m+n+1}...1_{n+1}0_n1_e...0_m...1_0}\Big)\\
\xrightarrow[]{CX(q_{e+n+1},q_{m+n+1}), CX(q_{e+n+1},q_{e}), CX(q_{e+n+1},q_{m})} \\
\frac{1}{\sqrt{2}} \Big(\ket{0_{2n+1}0_{e+n+1}...1_{m+n+1}...1_{n+1}0_n1_e...0_m...1_0}\\
+\ket{0_{2n+1}1_{e+n+1}...0_{m+n+1}...1_{n+1}0_n0_e...1_m...1_0}\Big)
\end{multline}
One may note that this requires only a couple of NOT (X) gates, one Hadamard (H) 
gate and a series of three CNOT (CX) gates.
\begin{figure*}[t]
    \centering
    \includegraphics[width= 17.25cm, height=7cm]{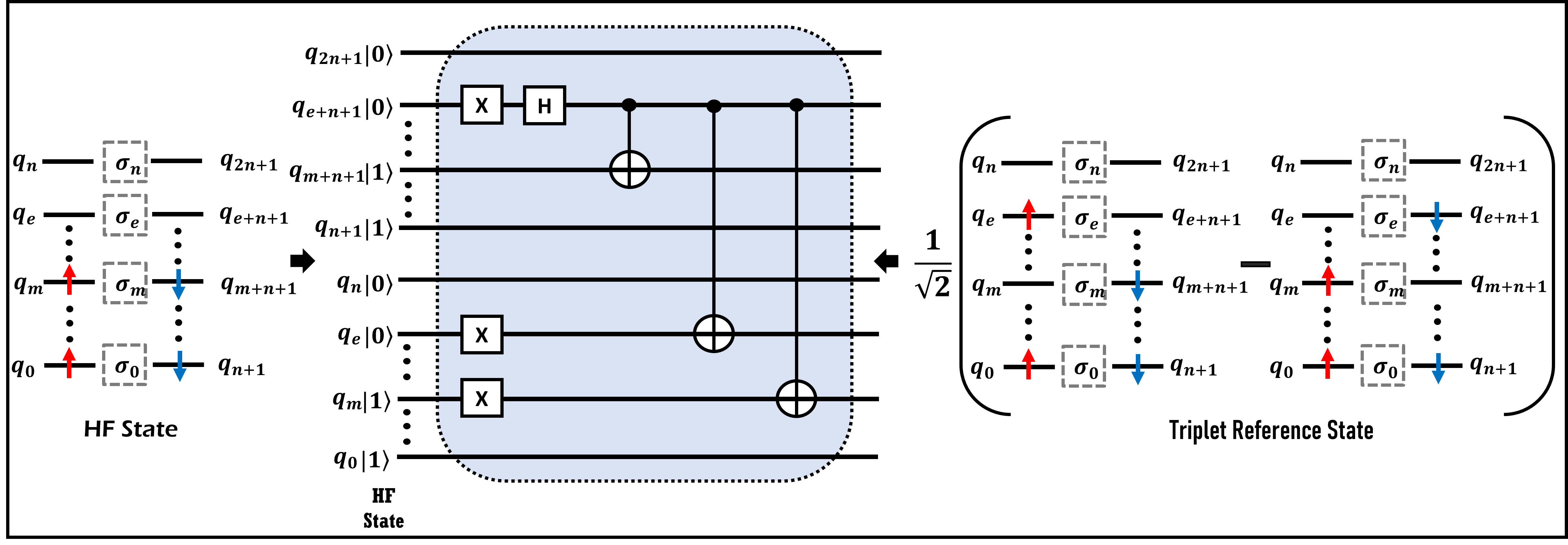}
    \caption{\textbf{Schematic quantum circuit to generate the triplet coupled reference with 
    spatial symmetry adapted to $\Sigma=\sigma_m\otimes\sigma_e$.}}
    \label{fig:circuit2}
\end{figure*}
Similarly, for the triplet reference state of symmetry $\Sigma$, one needs 
to construct a circuit for the entangled state: $\frac{1}{\sqrt{2}}(\ket{0_{2n+1}0_{e+n+1}...1_{m+n+1}...1_{n+1}0_n1_e...0_m...1_0}-\ket{0_{2n+1}1_{e+n+1}...0_{m+n+1}...1_{n+1}0_n0_e...1_m...1_0})$. This is 
shown is Fig. \ref{fig:circuit2}, which is somewhat similar to the previous 
circuit described previously, except the employment of an additional 
NOT ($X$) gate. The action of the sequence of the gates on the HF state is \
as follows:   
\begin{multline}
\ket{0_{2n+1}0_{e+n+1}...1_{m+n+1}...1_{n+1}0_n0_e...1_m...1_0}\xrightarrow[X(q_{m})]{X(q_{e+n+1}), X(q_e)} \\
\ket{0_{2n+1}1_{e+n+1}...1_{m+n+1}...1_{n+1}0_n1_e...0_m...1_0} \xrightarrow[]{H(q_{e+n+1})} \\
\frac{1}{\sqrt{2}}\Big(\ket{0_{2n+1}0_{e+n+1}...1_{m+n+1}...1_{n+1}0_n1_e...0_m...1_0}-\\
\ket{0_{2n+1}1_{e+n+1}...1_{m+n+1}...1_{n+1}0_n1_e...0_m...1_0}\Big)\\
\xrightarrow[]{CX(q_{e+n+1},q_{m+n+1}), CX(q_{e+n+1},q_{e}), CX(q_{e+n+1},q_{m})} \\
\frac{1}{\sqrt{2}} \Big(\ket{0_{2n+1}0_{e+n+1}...1_{m+n+1}...1_{n+1}0_n1_e...0_m...1_0}\\
-\ket{0_{2n+1}1_{e+n+1}...0_{m+n+1}...1_{n+1}0_n0_e...1_m...1_0}\Big)
\end{multline}
With this discussion on the reference state preparation that is adapted to
a given spatial and spin-symmetry, we now proceed to design the unitary ansatz
in terms of the totally symmetric spin-free operators.

\subsubsection{Choice of the Cluster Operators in terms of Spin-Free Unitary Generators:}\label{unitary_evo.}

We begin this section by noting that the conceptualization of the parametrization that 
enters in the unitary is nontrivial as our reference determinant now turns out to be 
a multi-determinantal contracted function rather than a single determinant. These determinants
have at least one electron in a spinorbital which was vacant in the HF determinant and at 
least one vacancy in the spinorbital which was occupied in the HF determinant. This makes
the choice of a physical vacuum to define the holes and particles somewhat ill-defined. 
In this section, we thus deliberately avoid using the term "hole" and "particle" orbitals,
rather we simply call them as occupied and unoccupied which are specific to 
the determinant under consideration. 

One may note that the determinants $\phi_A^\Sigma$ and $\phi_B^\Sigma$ are 
related to each other by a (set of) spin-complimentary excitation(s). 
The spin-complementary determinants may be defined as those 
related to each other where all the spin-states (up or down spin) of singly
occupied spatial orbitals are simultaneously interchanged. Thus, if one defines a 
spin-complementarity operator 
$\mathcal{P}(\bar{\phi}_m,{\phi}_m;{\phi}_e,\bar{\phi}_e)$ that simultaneously 
exchanges $\bar{\phi}_m$ by ${\phi}_m$ and ${\phi}_e$ by $\bar{\phi}_e$, it is
easy to note that 
$\mathcal{P}(\bar{\phi}_m,\phi_m;\phi_e,\bar{\phi}_e)\phi_A^\Sigma = \phi_B^\Sigma$.
Furthermore, since $\phi_A^\Sigma$ and $\phi_B^\Sigma$ have different 
occupancies in their spinorbitals, one may define determinant-selective 
(anti-hermitian) cluster operators, $\tau_A$ and $\tau_B$ which, by definition, 
act on $\phi_A^\Sigma$ and $\phi_B^\Sigma$ selectively. The cluster 
operators $\tau_C, C\in A, B$ are labelled by the parameters $\theta_C$ and the
string of creation/annihilation operators $\hat{\kappa}_C$.
Since the reference determinants are adapted to a given spin-multiplet, 
it is imperative that the relative weightage and phase of the correlated 
states remain conserved. This implies if the reference state is 
adapted to a given multiplet $S$ such that 
$\phi_0^{\Sigma, S}=\frac{1}{\sqrt{2}}(\phi_A^{\Sigma}\pm\phi_B^{\Sigma})$, 
one must be able to write the correlated state as: 
\begin{equation}
\psi^{\Sigma, S} = e^{\hat{\kappa}\theta}\frac{1}{\sqrt{2}}\Big(\phi_A^{\Sigma}\pm \phi_B^{\Sigma}\Big) 
\end{equation}
where the relative sign and factor remain the same as the multi-determinantal 
reference. 
Depending on the structure of
the operator string, $\tau$ can have two components: 
$\tau = \tau_A \cup \tau_B$.
$\tau_A$ that acts non-trivially on the determinant 
$\phi_A^\Sigma$ while it \textit{may or may not} have a 
non-vanishing action on $\phi_B^\Sigma$ and 
$\tau_B$ which is defined otherwise. This means 
there may be some operators $\tau$ which act on
both the determinants. In order to maintain the spin-
adaptation of the wavefunction throughout the
optimization process, various excited determinants 
of the evolved
wavefunction that are related to each other by 
spin-complimentary terms must be treated in the 
same footings. This mandates one to impose 
additional restrictions on certain
spin-complimentary excitations of $\tau_A$ and 
$\tau_B$ rather than leaving them as free 
variational parameter.
This can be achieved by equating some of the
spin-complimentary excitation amplitudes $\theta_A$ and $\theta_B$ as shown below.
The general philosophy of such restriction is that if a certain excitation  
(generically labeled as $I$) of $\hat{\kappa}_A$ generates a
determinant by its action on $\phi_A^{\Sigma}$ which is spin-complementary 
to the determinant generated by the $J$-th excitation of $\hat{\kappa}_B$ 
on $\phi_B^{\Sigma}$, then $\theta_{A_I}$ and $\theta_{B_J}$
are taken to be equal. Mathematically, if
\begin{equation}
    \mathcal{P}(..,..;..,..)\Bigg(\hat{\kappa}_{A_I}\ket{\phi_A^{\Sigma}}\Bigg) = \hat{\kappa}_{B_J}\ket{\phi_B^{\Sigma}} 
    \label{spincomp}
\end{equation}
then, $\theta_{A_I}=\theta_{B_J}$.
Let us discuss this with a few cases of a more concrete example with one-
and two-body operators.
While the two-body amplitudes are also chosen in a way that satisfies Eq. \ref{spincomp}, 
we do not show all these cases explicitly here since there are numerous possibilities.

\vspace{0.05cm}
\noindent \textbf{\textit{One-body operators:}}

\textbf{Case I}: Let us consider a determinant where the excitations are from a 
core occupied orbital ($\phi_i$) to high-lying unoccupied orbital ($\phi_a$). This 
assumes that both the up and down-spinorbitals corresponding to $\phi_i$ are 
occupied and contrarily, both the up and down-spinorbitals corresponding to $\phi_a$ 
are unoccupied in both $\phi_A^\Sigma$ and $\phi_B^\Sigma$. Let us consider the 
action of the operators $\hat{\kappa}_{i}^{a}$ and $\hat{\kappa}_{\bar{i}}^{\bar{a}}$ on the multi-determinantal reference. 
Since both the operators act upon both the determinants, for this particular example,
we drop the subscript $A$ or $B$ from the operators.
\begin{eqnarray}
\hat{\kappa}_{i}^{a}\frac{1}{\sqrt{2}}(\ket{\phi_A^{\Sigma}}\pm\ket{\phi_B^{\Sigma}}) = \frac{1}{\sqrt{2}}(\ket{\phi_{A_{i}}^{\Sigma_{a}}}\pm\ket{\phi_{B_{i}}^{\Sigma_{a}}}) \nonumber \\
\hat{\kappa}_{\bar{i}}^{\bar{a}}\frac{1}{\sqrt{2}}(\ket{\phi_A^{\Sigma}}\pm\ket{\phi_B^{\Sigma}}) = \frac{1}{\sqrt{2}}(\ket{\phi_{A_{\bar{i}}}^{\Sigma_{\bar{a}}}}\pm\ket{\phi_{B_{\bar{i}}}^{\Sigma_{\bar{a}}}})
\end{eqnarray}
At this stage, one may note that
\begin{eqnarray}
\mathcal{P}(\bar{\phi}_{i},{\phi}_{i};{\phi}_{a},\bar{\phi}_{a})\ket{\phi_{A_{i}}^{\Sigma_{a}}}=\ket{\phi_{B_{\bar{i}}}^{\Sigma_{\bar{a}}}} \nonumber \\
\mathcal{P}(\bar{\phi}_{i},{\phi}_{i};{\phi}_{a},\bar{\phi}_{a})\ket{\phi_{B_{i}}^{\Sigma_{a}}}=\ket{\phi_{A_{\bar{i}}}^{\Sigma_{\bar{a}}}}
\end{eqnarray}
In order to preserve the spin-symmetry of the excited determinants thus generated, 
one must restrict the variational freedom by imposing the condition.
\begin{equation}
     \theta_{i}^{a}=\theta_{\bar{i}}^{\bar{a}}    
\end{equation}

\textbf{Case II}: We now consider the situation where an operator like $(\hat{\kappa}_A)_{\bar{m}}^{\bar{a}}$ has a non-vanishing action only on $\phi_A^{\Sigma}$ while $(\hat{\kappa}_B)_m^a$ has a non-vanishing action only on $\phi_B^{\Sigma}$. We now analyze the action of operator $(\hat{\kappa}_A)_{\bar{m}}^{\bar{a}}$ and
$(\hat{\kappa}_B)_{m}^{a}$
on the multi-determinantal reference state.
\begin{eqnarray}
    (\hat{\kappa}_A)_{\bar{m}}^{\bar{a}}\frac{1}{\sqrt{2}}(\ket{\phi_A^{\Sigma}}\pm\ket{\phi_B^{\Sigma}}) = \frac{1}{\sqrt{2}}(\ket{\phi_{A_{\bar{m}}}^{\Sigma_{\bar{a}}}}) \nonumber \\
(\hat{\kappa}_B)_{m}^{a}\frac{1}{\sqrt{2}}(\ket{\phi_A^{\Sigma}}\pm\ket{\phi_B^{\Sigma}}) = \frac{1}{\sqrt{2}}(\pm\ket{\phi_{B_{m}}^{\Sigma_{a}}})
\end{eqnarray}
One may however note that:
\begin{eqnarray}
   \mathcal{P}(\bar{\phi}_{a},{\phi}_{a};{\phi}_{e},\bar{\phi}_{e})\ket{\phi_{A_{\bar{m}}}^{\Sigma_{\bar{a}}}}=\ket{\phi_{B_{m}}^{\Sigma_{a}}}
\end{eqnarray}
Again, to preserve the spin-symmetry of the evolved state, one must ensure that:
\begin{equation}
    (\theta_A)_{\bar{m}}^{\bar{a}}=(\theta_B)_{m}^{a}
\end{equation}

One may generalize the above conditions for double excitation operators. We will not derive all the cases separately but rather give the final working 
condition explicitly that must be satisfied to 
conserve the reference spin-symmetry.

\begin{figure*}[t]
    \centering
    \includegraphics[width= 18cm, height=12cm]{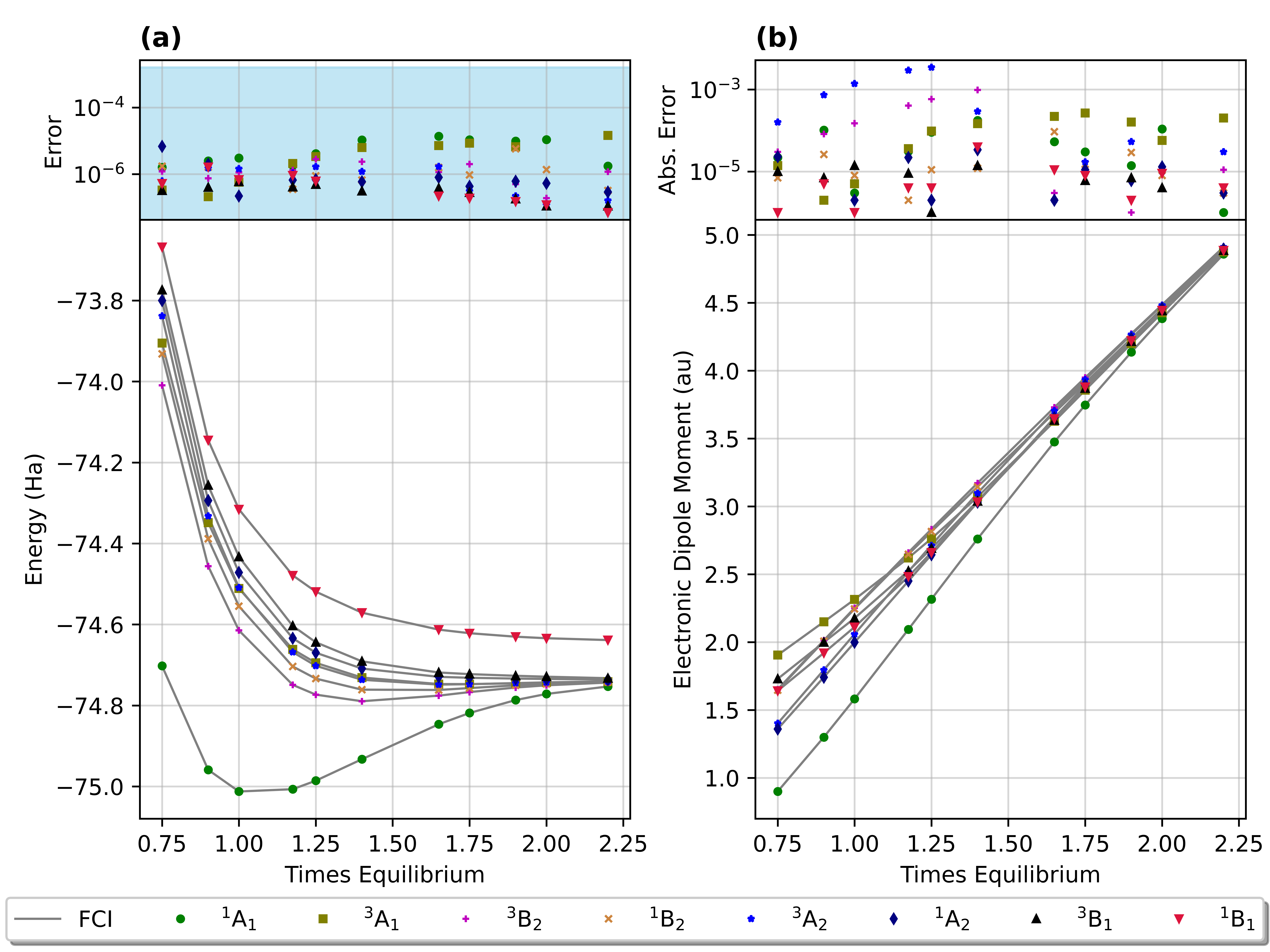}
    \caption{\textbf{The potential energy profile (a) and the electronic dipole moment surface (b) for the ground and various excited states of $H_2O$ along its symmetric stretching pathway. The solid lines are the FCI values whereas the dots are the results obtained from state-specific VQE (ss-VQE). Clearly, they are in excellent agreement. The top panels show the difference in the magnitude of the quantities from FCI values.}}
    \label{fig:h2o}
\end{figure*}
\begin{figure*}[t]
    \centering
    \includegraphics[width= 18cm, height=10cm]{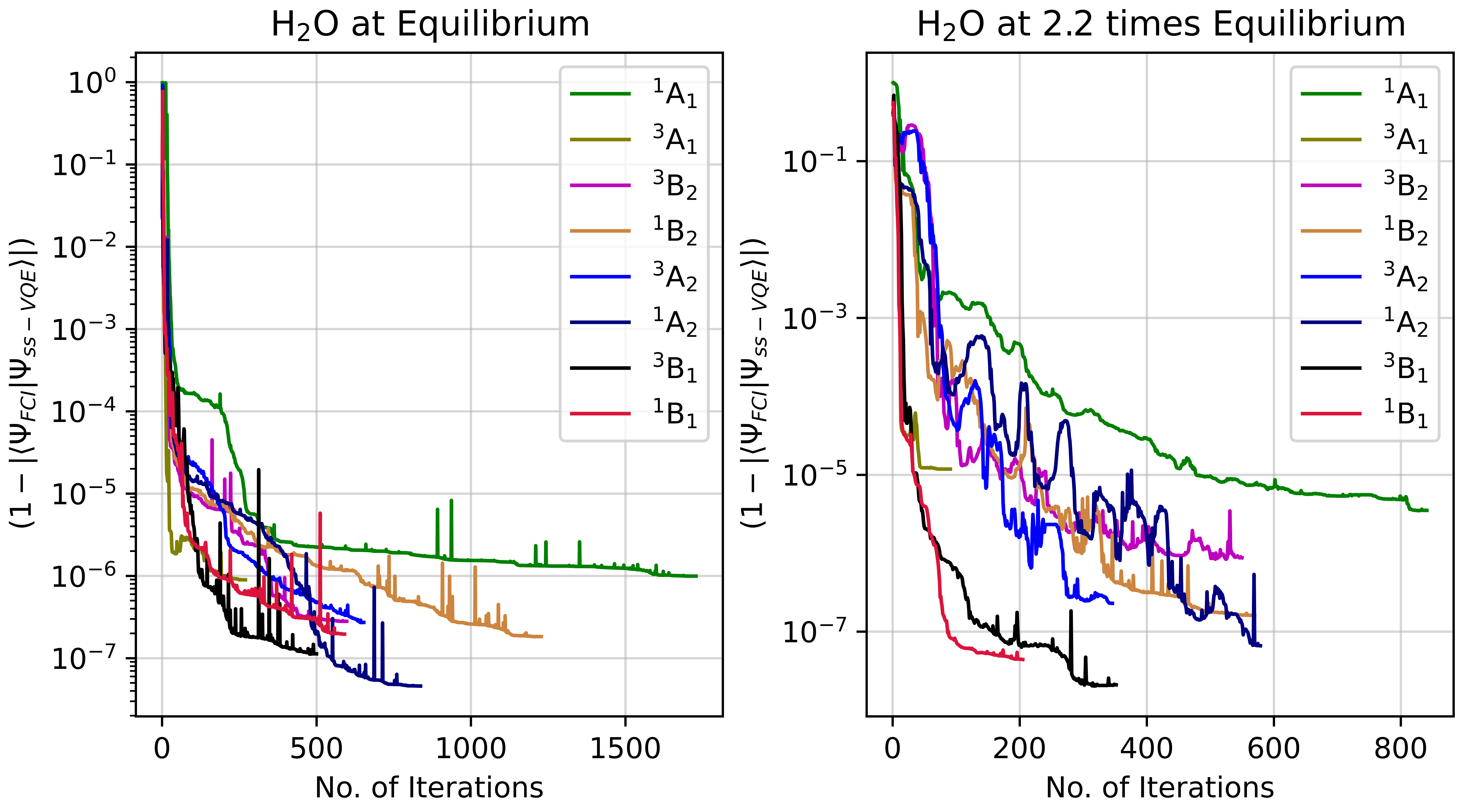}
    \caption{\textbf{Difference in overlap for various singlet and triplet roots determined via ss-VQE with the corresponding Full Configuration Interaction wavefunction for two different geometries of $H_2O$.}}
    \label{fig:overlap}
\end{figure*}

\vspace{0.05cm}
\noindent \textbf{\textit{Two-body operators:}}

\textbf{Case I:} For the excitations from 
core orbitals (doubly occupied in both $\phi_A^\Sigma$ and $\phi_B^\Sigma$) to orbitals which are unoccupied in both $\phi_A^\Sigma$ and $\phi_B^\Sigma$. These are the cases where
both $\hat{\kappa}_A$ and $\hat{\kappa}_B$ have non-vanishing action on $\phi_A^\Sigma$ and $\phi_B^\Sigma$.
\begin{eqnarray}
    (\theta)_{i,\bar{j}}^{a,\bar{b}}=(\theta)_{\bar{i},j}^{\bar{a},b} \nonumber \\
    (\theta)_{i,j}^{a,b}=(\theta)_{\bar{i},\bar{j}}^{\bar{a},\bar{b}}
\end{eqnarray}
\textbf{Case II:} These are the cases where $\hat{\kappa}_A$ has a non-vanishing action only on $\phi_A^\Sigma$ and $\hat{\kappa}_B$ has a non-vanishing action only on $\phi_B^\Sigma$. To cite an example, one may choose
operator strings like $(\hat{\kappa})_{\bar{m},i}^{\bar{e},a}$ and $(\hat{\kappa})_{m,\bar{i}}^{e,\bar{a}}$. To ensure that the
resulting determinants generated by the action of these operators are spin-adapted, one must ensure that
\begin{equation}
     (\theta_A)_{\bar{m},i}^{\bar{e},a}=(\theta_B)_{m,\bar{i}}^{e,\bar{a}}
\end{equation}
A similar analysis shows that
\begin{equation}
    (\theta_A)_{\bar{m},\bar{i}}^{\bar{e},\bar{a}}=(\theta_B)_{m,i}^{e,a}
\end{equation}
\textbf{Case III:} These are the cases for which involve paired excitations ($\hat{\kappa}_p$). 
By definition, they have a non-vanishing action on both $\phi_A^\Sigma$ and $\phi_B^\Sigma$ and thus 
this case may be assumed to be a sub-class of case-I. 

The final ansatz towards a state-selective evolution of the reference state is 
constructed in a disentangled manner with the restrictions on the spin complementary 
excitation amplitudes. Since our purpose of this manuscript is to demonstrate 
that such an ansatz may lead to a targetted excited state in a state-selective 
manner, we have not considered its optimal ordering to fine-tune the accuracy. 
Furthermore, no parameter thresholding criteria have been employed to 
restrict the ansatz depth. Rather, we have taken a lexical ordering of
the operators to construct the factorized unitary where all the one 
body operators act first on the multi-determinantal reference, followed by 
the two-body operators in a lexical order. With the various class of excitations
that enter into the parametrized unitary, one may write the final ansatz as 
\begin{eqnarray}
\begin{aligned}
    U(\theta)& =e^{\big((\hat{\kappa}_p)_{i,\bar{i}}^{a,\bar{a}}\big)\theta_7}...e^{\big((\hat{\kappa}_A)_{\bar{m},\bar{i}}^{\bar{e},\bar{a}}+(\hat{\kappa}_B)_{m,i}^{e,a}\big)\theta_6}...\\
    & e^{\big((\hat{\kappa})_{i,j}^{a,b}+(\hat{\kappa})_{\bar{i},\bar{j}}^{\bar{a},\bar{b}}\big)\theta_5}...e^{\big((\hat{\kappa}_A)_{\bar{m},{i}}^{\bar{e},{a}}+(\hat{\kappa}_B)_{{m},\bar{i}}^{{e},\bar{a}}\big)\theta_4}\\
    & ...e^{\big((\hat{\kappa})_{i,\bar{j}}^{a,\bar{b}}+(\hat{\kappa})_{\bar{i},j}^{\bar{a},b}\big)\theta_3}
    ...e^{\big((\hat{\kappa}_A)_{\bar{m}}^{\bar{a}}+(\hat{\kappa}_B)_{{m}}^{{a}}\big)\theta_2}...e^{\big(\hat{\kappa}_{i}^{a}+\hat{\kappa}_{\bar{i}}^{\bar{a}}\big)\theta_1}
\end{aligned}    
\end{eqnarray}
Note that in writing so, we have only shown how various operator classes are groupped; however, in writing so,
it is difficult to account for the lexicographic ordering.
In the actual implementation, the operators are ordered in their lexical ordering.

As we discussed previously, the restriction on the 
parametrization renders the unitary defined above to be 
a spin scalar. Since the Hamiltonian is also space-spin 
scalar, it is easy to see that the variational optimization
starting from a reference that is adapted to a spatial
symmetry $\Sigma$ and spin-multiplet $S$ would land up
in the lowest energy eigenroot which also has the
same space-spin symmetry. 
\begin{equation}
E_0^{\Sigma,S}=\bra{\phi_0^{\Sigma,S}}U^\dagger(\theta)HU(\theta)\ket{\phi_0^{\Sigma,S}}
\end{equation}
where $E_0^{\Sigma,S}$ is the lowest eigen-energy of the 
state specified by the spatial symmetry $\Sigma$ and 
spin-symmetry $S$. One may note that such a state, except 
when $\Sigma$ is totally symmetric and singlet, the various 
roots (characterized by different combinations of 
$\Sigma$ and $S$) are the spectra of low-lying excited
states. Within a variational eigensolver 
parlance, the action of the space-spin scalar 
unitary on the symmetry adaptation of the reference
brings in the state-specificity of the solution where each 
root can be selectively targeted. 

It should further be noted that the choice of the 
multi-determinantal reference may often lead to a redundant 
description of the excited determinant. For example,
$(\hat{\kappa}_A)_{e,\bar{m}}^{a,\bar{b}}\ket{\phi_A^\Sigma}$ and 
$(\hat{\kappa}_B)_{m,\bar{e}}^{a,\bar{b}}\ket{\phi_B^\Sigma}$
lead to the same excited determinant. Similarly, their 
spin-complimentary excitations also generate the same 
excited determinant. While each of these excited determinants
should be generated once, we assign
$(\theta_A)_{e,\bar{m}}^{a,\bar{b}}$ and 
$(\theta_B)_{m,\bar{e}}^{a,\bar{b}}$ as different variational 
parameters. Similarly, their spin-complementary counterparts
$(\theta_B)_{m,\bar{e}}^{b,\bar{a}}$ and 
$(\theta_A)_{e,\bar{m}}^{b,\bar{a}}$ are also treated as two
different parameters. Moreover, cases may appear where 
the action of a one-body excitation and a two-body 
excitation leads to the same excited determinant. This 
redundant parametrization may plague the variational
optimization trajectory. However, in the systems we have studied,
we have not encountered such numerical instability as the variational flexibility has navigated past any such problem that
might have appeared. One may note that this redundant 
parametrization is not in contradiction to Eq. \ref{spincomp}
as $(\theta_A)_{e,\bar{m}}^{a,\bar{b}}$ and
$(\theta_B)_{m,\bar{e}}^{b,\bar{a}}$ (and similarly 
$(\theta_B)_{m,\bar{e}}^{a,\bar{b}}$ and $(\theta_A)_{e,\bar{m}}^{b,\bar{a}}$) are taken to be equal.

One may further note that the determination of the low-lying
excited roots does not necessitate the knowledge of the 
ground state solution. This is a clear advantage over 
q-sc-EOM\cite{eom_vqe} or qEOM\cite{qEOM}, where the excited state description is 
built on top of the correlated ground state solution. The
sequential approach towards the determination of the excited
state energetics renders to be more impacted by the 
NISQ device noise. It is trivial to note that our 
state-specific approach introduces only as much error
as any other ground state calculation and is hence expected
to be more accurate. Furthermore, the existing methods 
to determine excited states often impose additional
orthogonality constraints or some additional qubits 
in order to maintain orthogonality with the previously
calculated lower roots. At least for the low-lying 
excited energy spectrum, our method bypasses the 
requirement of orthogonality constraint. However, one 
requires to impose such orthogonality constraint only
when one targets the second lowest root of a given spatial
and spin symmetry.

\section{Results and Discussions}\label{results}
\subsection{Methodology:}
In all our implementations, we have used qiskit-nature\cite{Qiskit} as an 
interface medium that imports the one- and two-body integrals and also the 
symmetry of each spinorbital from PySCF\cite{pyscf}. We have used STO-3G basis set\cite{sto} in 
all the applications studied in this manuscript. The spinorbitals are directly 
mapped to qubits. In order to transform the second quantized operators into qubit 
operators we have used Jordan-Wigner encoding\cite{JWT}. For all the calculations that we 
have performed, we have chosen the statevector-simulator assuming an ideal quantum 
environment, and L-BFGS-B optimizer\cite{lbfgs} was used for the classical optimization of
the amplitudes.

In all our applications, we provide more-than-required variational
flexibility to our ansatz by repeating the entire circuit
twice. While for some cases, this may artificially make the predicted energy 
more accurate at the cost of increase in the gate depth or
the parameter count that may go beyond the NISQ realization,
we do so for a sanity-check: As a proof of the concept, we
demonstrate that somewhat over-parametrization in the ansatz
does not guide the optimized energy below the targetted state
towards the \textit{absolute} ground state. Since the number 
of free parameters are restricted in order to maintain spin-symmetry,
even the doubling of the circuit leads to a very compact structure 
of the ansatz in terms of only a few variational parameters, while the 
gate depth is comparable to the conventional UCCSD.
We provide below numerical justifications for a number of challenging
systems to justify our claim.

\subsection{Symmetric stretching of $H_2O$}
The symmetric stretching of $H_2O$ is one of the most well-referred test cases
to study the interplay of various degrees of correlation effects. 
In STO-3G basis, it has 14 spin-orbitals. 
In our calculation we have frozen the core 1s orbital of $O$, making it a case of
8 electrons in 12 spinorbitals. We have fixed the $H-O-H$ bond angles at $104.4776^{\circ}$
while symmetrically stretched both the $O-H$ bonds, keeping the molecular symmetry 
of $H_2O$ to be $C_{2v}$ throughout the pathway. $C_{2v}$ point group has 
4 irreducible representations: $A_1, A_2, B_1, B_2$. Our state-specific VQE 
is employed to directly determine the 
lowest singlet and triplet roots that are characterized by these irreducible 
representations. One may note that these roots are some of the low-lying
excited states and a state-specific determination of these roots are over
the potential energy surface is extremely challenging. The FCI calculations 
suggest that the $^1A_1$ state has an attractive surface with distinct 
minima, while $^3A_1, ^3B_2, ^1B_2$ and $^3A_2$ show very shallow minima. Furthermore, except
$^1B_1$, all the roots have the same asymptotically converged energy value. The 
state-specific VQE with a two-fold repetition of the circuit shows
very high accuracy with all the roots (Fig. \ref{fig:h2o}a). More importantly, 
none of these roots fall below the targetted value, justifying the importance 
of state-specificity of the VQE parametrization. We must mention
that, as a proof of the concept, the ground state ($^1A_1$) calculations were 
performed taking a high-lying symmetry adapted and 
singlet-coupled multi-determinantal reference, rather than 
the usual HF determinant. With doubling the circuit, the 
variational flexibility, however, leads the optimization 
towards the exact ground state energy values. One may of course 
choose to start with the HF reference to determine the singlet $A_1$ root and in that case, the
doubling of the circuit will not be required for accuracy. For 
all other roots, our choice of the reference was energetically
optimal.
\begin{figure*}
    \includegraphics[width=10.5 cm,height=11.5 cm]{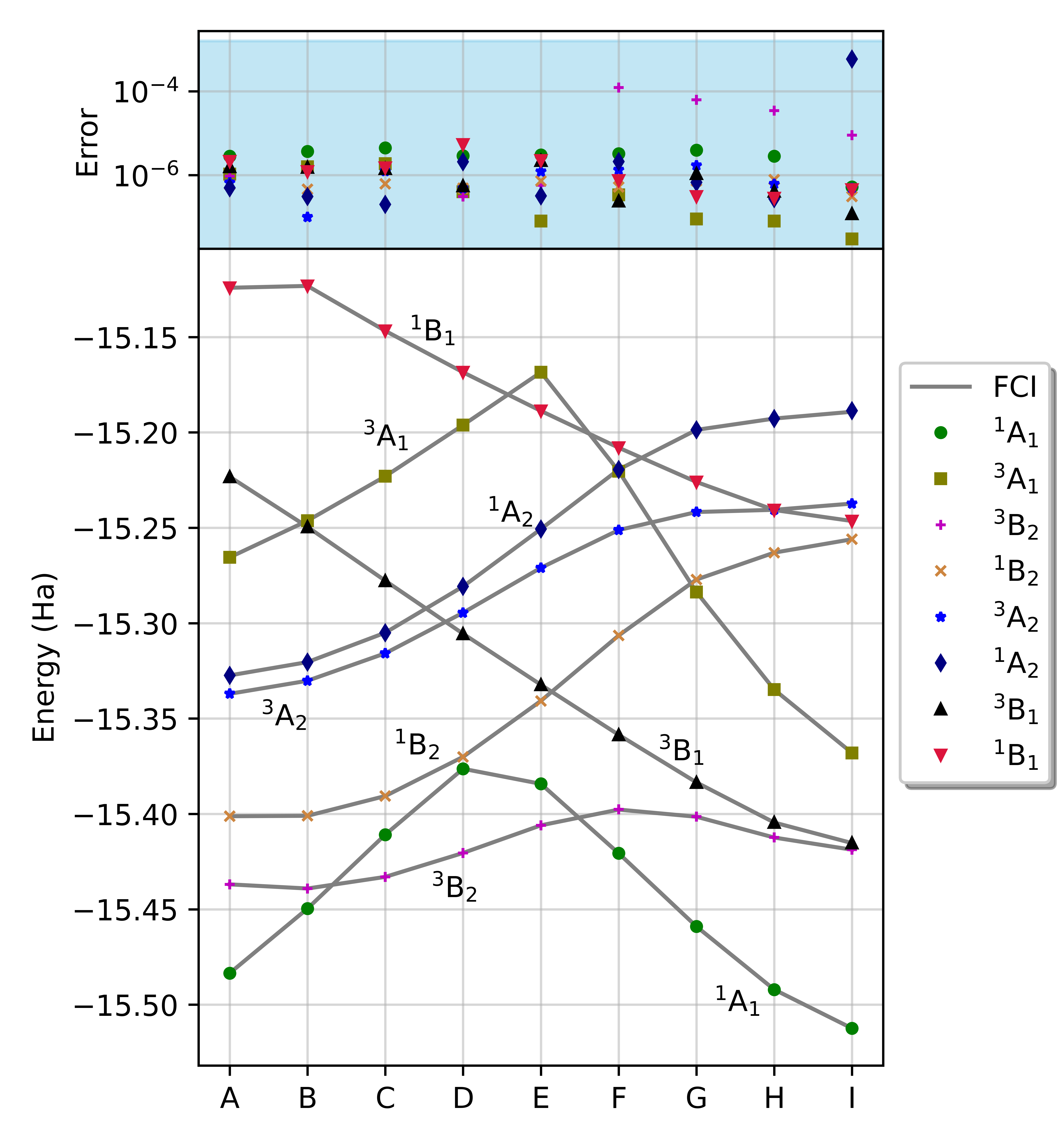}
    \caption{\textbf{Potential energy profile for ground and various excited states along the reaction path of perpendicular insertion of $Be$ in $H_2$. The Solid lines are the FCI values whereas the dots are the results obtained from state-specific VQE (ss-VQE)}}\label{fig:beh2}
  \end{figure*}

We have calculated the electronic dipole moment at each geometry along the 
$C_{2v}$ stretching pathway and compared it with the FCI values. The dipole
moment surface for each state with state-specific VQE is also shown 
in fig\ref{fig:h2o}b and clearly they are in excellent agreement with the 
FCI values. Additionally, we have also calculated the overlap of the 
wavefunction for each of these ground and excited states with the FCI counterparts at 
each iterative step of the optimization process (Fig. \ref{fig:overlap}). The 
high accuracy in the determination of the wavefunctions for each irreducible representation,
as demonstrated by their overlap with the FCI counterparts justifies the 
state-specificity of the optimization process towards excited state calculations.

\subsection{$C_{2v}$ insertion of Be in $H_2$}

\begin{table}[htbp]
    \centering
    \setlength{\tabcolsep}{12pt}
\renewcommand{\arraystretch}{1.6}
 \begin{tabular}{cccc} 
 \hline\hline
 Point & X & Y & Z \\ [0.5ex] 
 \hline
 A & 0.0 & $\pm$1.62 & 2.00 \\ 
  B & 0.0 & $\pm$1.505 & 2.25 \\
  C & 0.0 & $\pm$1.39 & 2.50 \\
  D & 0.0 & $\pm$1.275 & 2.75 \\
  E & 0.0 & $\pm$1.16 & 3.00  \\  
  F & 0.0 & $\pm$1.045 & 3.25 \\ 
   G & 0.0 & $\pm$0.93 & 3.50 \\
   H & 0.0 & $\pm$0.815 & 3.75 \\
   I & 0.0 & $\pm$0.70 & 4.00 \\
   \hline\hline
\end{tabular}
\caption{Cartesian coordinates of H atoms (in Units of Bohr)}
\label{Table:1}
\end{table}


The perpendicular insertion of $Be$ atom in $H_2$ is one of the most challenging test cases 
for any newly developed quantum mechanical theory. The insertion pathway may be characterized
by several discrete points as shown in Tab. \ref{Table:1} as suggested by Purvis \textit{et. al.}\cite{purvis} where 
the $Be$ atom is placed at $(0.0,0.0,0.0)$. Around points C, D, and E, the system 
shows strong multi-reference character and thus it becomes extremely difficult for any 
single reference theory to achieve quantitative accuracy. We simulated the ground and
a number of excited states along the insertion pathway in STO-3G basis where the core
$1s$ orbital of $Be$ is kept frozen. This renders the system to consist of 4 electrons
in 12 spinorbitals. We would like to point out two remarkable observations: The $^1A_1$
root can be obtained with a high degree of precision when started from a high-lying multi-determinantal 
singlet $A_1$ reference with the doubling of the circuit. One may note that with the 
restriction of the spin-complementary excitations, the doubling of the circuit 
does not lead to an increase in the number of free parameters beyond the usual UCCSD. It was 
previously observed by Sugisaki \textit{et. al.}\cite{sugisaki_beh2} that one may achieve this level of precision when
started from multi-configurational SCF orbitals with generalized pair excitations. 
However, when one starts either from the HF or the MCSCF wavefunction, any arbitrary UCC
parametrization will only lead to the lowest $^1A_1$ state and miss out on any lower
energy state that may exist at some points of the energy profile. It was
observed by Sugisaki \textit{et. al.}\cite{sugisaki_beh2} that the classical FCI calculations reveal the existence of 
$^3B_2$ root energetically below $^1A_1$ at points $C, D$ and $E$. While the 
conventional UCC with HF or MCSCF reference is unable to detect this, the ss-VQE, 
when started from the $^3B_2$ multi-determinantal reference can state-selectively
lead to the corresponding eigenstates. Also, few other roots, particularly $^3A_1$ 
shows the signature of a strong correlation around points D and E as predicted by 
FCI calculations; the ss-VQE can capture each of such behavior for all 
the low-lying roots with correct quantitative and qualitative accuracy. The 
monotonic convergence to the FCI values for each root, in spite of the possible 
over-parametrization in the ansatz, validates the state-specificity of our optimization
scheme.

\section{Conclusions and Future Outlook:} \label{conclusion}
In this manuscript,
we introduced a unified variational quantum eigensolver 
framework that treats both ground and any low-lying 
excited state of arbitrary spatial and spin symmetry in 
exactly the same footing. In doing so, we have brought 
back the fundamental concepts of symmetry 
of electronic states that, although supremely important, 
had hitherto remained fuzzy in quantum computing. 
The approach is pivoted upon the construction of a 
multi-determinantal reference that is characterized by 
the irreducible representation of the target state and 
a desired spin multiplicity. The reference is adapted to 
the given spin-multiplet by entangling its determinants
appropriately through Clebsch-Gordan coefficients. The 
parametrization in the unitary is chosen to be totally 
symmetric and in a certain spin-restricted manner that
maintains spin-scalarity. Our approach is state-specific
in nature: it targets one root (with a given irreducible 
representation and spin symmetry) at a time as desired and
does not require any prior knowledge of the ground state.
The method scales exactly the same for both ground and 
excited states as dictated by the parametrization of the
ansatz. In fact, its realization in a quantum architecture 
requires fewer free parameters than an unrestricted
variational protocol due to the spin-complementarity
criterion. Due to the direct access to the 
excited states, it remains unaffected from any cumulative 
error that is expected to alter the accuracy of any excited
state calculations. Furthermore, it requires no additional
qubit or post-processing to restore spin symmetry. 
One may further trivially couple it with variational 
quantum deflation to access higher excited states. While 
this work paves the foundation to a direct, resource-efficient
and resilient approach to the excited state, one may further 
aim to compactify the ansatz for excited states via 
approaches like ADAPT-VQE\cite{ADAPT} or COMPASS\cite{COMPASS}.

\section{Acknowledgement}
DM thanks Prime Minister's Research Fellowship (PMRF), Government of India for his research fellowship. The authors thank Mr. Anish Chakraborty for all the valuable discussions during the development of the theory.


\subsection*{DATA AVAILABILITY}
The numerical data that support the findings of this study are
available from the corresponding author upon 
reasonable request.

\bibliography{literature}
\end{document}